\documentclass[conference]{IEEEtran}
\IEEEoverridecommandlockouts
\usepackage{cite}
\usepackage{amsmath,amssymb,amsfonts}
\usepackage{algorithmic}
\usepackage{graphicx}
\usepackage{textcomp}
\usepackage{xcolor}
\def\BibTeX{{\rm B\kern-.05em{\sc i\kern-.025em b}\kern-.08em
    T\kern-.1667em\lower.7ex\hbox{E}\kern-.125emX}}
\usepackage{caption}
\usepackage{subcaption}

\begin{document}

\title{Rethinking Reliability Using Network Coding: a Practical 5G Evaluation\\
\thanks{This material is based upon work supported by JMA Wireless.}
}

\author{\IEEEauthorblockN{Laura Landon$^{\dagger}$, Vipindev Adat Vasudevan$^{\dagger}$, Junmo Sung$^{*}$, and Muriel M\'edard$^{\dagger}$}
\IEEEauthorblockA{
$^\dagger$Massachusetts Institute of Technology (MIT), Cambridge, USA, Emails: \{llandon9, vipindev, medard\}@mit.edu }
\IEEEauthorblockA{
$^*$JMA Wireless, Syracuse, USA, Email: \{jsung\}@jmawireless.com }
}

\maketitle
\begin{figure}[b]
\small
\copyright 2025 IEEE.  Personal use is permitted.  Permission required for all other uses, in any current or future media, including
reprinting/republishing this material for advertising or promotional purposes, creating new collective works,
for resale or redistribution to servers or lists, or reuse of any copyrighted component of this work in other works. This document is accepted to be published in Proc. IEEE 50th Conf. on Local Computer Networks (LCN), Oct. 2025.
\end{figure}

\begin{abstract}
    This work presents the design and implementation of a real-time network coding system integrated into the IP layer of a 5G testbed, offering an alternative to conventional retransmission-based reliability mechanisms such as ARQ and HARQ. Using a netfilter-based packet interception framework, we inject forward erasure correction using Random Linear Network Coding (RLNC) into live traffic between a gNB and UE over a 3GPP RF link. We evaluate a block coding scheme, analyzing its impact on throughput, jitter, and resource usage. Results show that with appropriate code rate selection, RLNC can fully recover from packet losses using fewer transmissions than ARQ/HARQ and maintain a high throughput, particularly under moderate-to-high packet loss rates. These findings demonstrate that network coding can effectively replace retransmission-based reliability in future wireless systems, with the potential for more efficient resource utilization.
\end{abstract}

\begin{IEEEkeywords}
random linear network coding (RLNC), HARQ, ARQ, 5G mobile communication, network reliability
\end{IEEEkeywords}

\section{Introduction}

\begin{figure*}
    \centering
    \includegraphics[width=\textwidth]{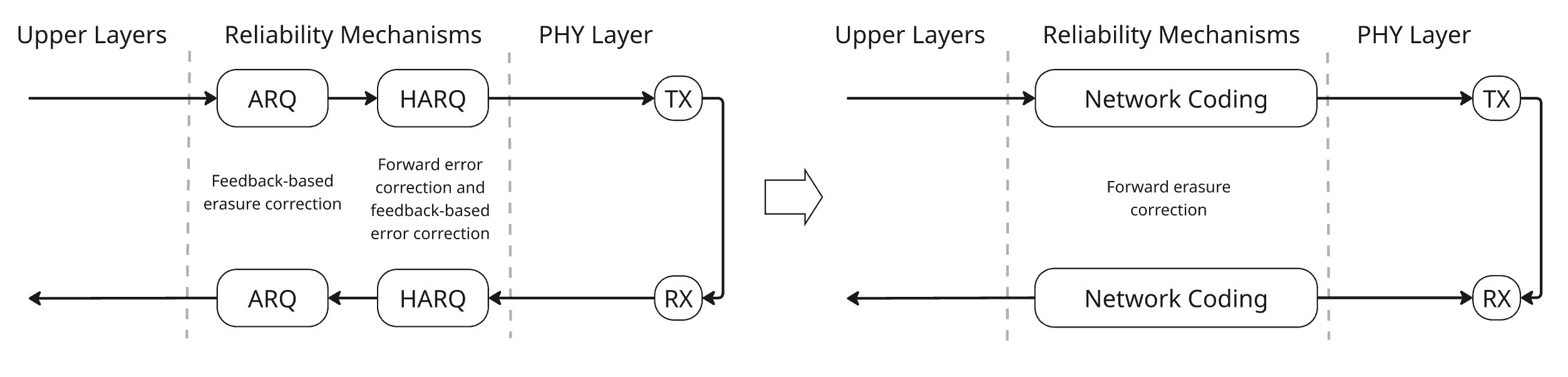}
    \caption{In the current 5G protocol stack, error correction is handled through HARQ in the MAC layer and ARQ in the RLC layer, both relying on feedback. By using a forward erasure correction (FEC) method like network coding, it is possible to reduce the delay in error correction. Replacing these layered, feedback-dependent error correction mechanisms with FEC could significantly lower latency in future wireless systems.}
    \label{fig:intro_figure}
\end{figure*}

Mobile networks are continuously evolving to support greater demand and stringent performance requirements. The International Telecommunication Union (ITU) has put forth the emerging trends for 6G \cite{wp5d2023m}, including the Hyper-Reliable Low-Latency Communication (HRLLC). Current cellular systems primarily rely on Automatic Repeat reQuest (ARQ) and Hybrid ARQ (HARQ) for reliability, but these retransmission-based schemes incur latency penalties proportional to the round-trip time (RTT). Faster erasure correction methods are required to meet the requirements of future wireless networks. In situations where latency is a priority, network coding offers an attractive alternative: send corrective packets in advance.


This paper explores Random Linear Network Coding (RLNC) \cite{ho2006random} as a proactive alternative to ARQ/HARQ within the 5G protocol stack. Rather than retransmitting lost packets, RLNC introduces forward erasure correction by transmitting algebraic combinations of packets as redundancy in advance. When the necessary number of redundant packets is correctly obtained, the receiver can recover the original data without requiring retransmissions, reducing packet service time on average and simplifying the system of error and erasure correction in 5G (Fig. \ref{fig:intro_figure}).

Previous work evaluating the relative benefit of RLNC and HARQ/ARQ has simulated RLNC at the MAC layer, the same layer at which HARQ is implemented \cite{landon2024enhancing}. However, implementing RLNC in the MAC layer on real systems requires modification of those 5G implementations, for which access is not always simple to obtain. An alternative method that does not require access to the lower layers is to implement network coding in the IP layer. This allows network coding to be tested and used more easily on existing systems. This paper draws on the work in \cite{teerapittayanon2012network} to create a network coding layer between the IP layer and the 3GPP protocol stack and reports performance results taken from an implementation of this layer on a gNodeB (gNB) and a user equipment (UE) connected by a 5G RF link. The experimental measurements over this realistic scenario for different transmission gains and code rates showcase the benefits of network coding. For a loss rate of 20\%, network coding achieves a performance improvement of up to 5.9 times compared to the existing HARQ/ARQ based approach.

The rest of this paper is organized as follows: Section II reviews prior work on URLLC and network coding. Section III describes the system design and implementation of our IP-layer RLNC framework, including the packet interception mechanism and coding modules. Section IV presents experimental results comparing RLNC with conventional HARQ/ARQ mechanisms in terms of throughput, jitter, and resource usage. Section V provides a discussion on the trade-offs between network coding and retransmission-based methods, with particular attention to redundancy and jitter behavior. Section VI outlines directions for future work, including sliding window decoding and integration into lower layers of the 5G protocol stack, and Section VII concludes the paper.

\section{Related Work}


Next-generation wireless systems are expected to support ultra-reliable low-latency communication (URLLC), which places stringent requirements on both reliability (e.g., packet error rates below $10^{-5}$) and latency (as low as 1 ms end-to-end) across diverse use cases ranging from industrial automation to remote surgery. Meeting these demands requires a combination of physical-layer advancements and architectural innovations \cite{alves2021beyond, pourkabirian2024vision}.

In 5G New Radio (NR), reliability is achieved through a combination of error correction and feedback-driven retransmission mechanisms spanning multiple protocol layers. The physical layer employs adaptive modulation and coding scheme (MCS) selection to target a given block error rate on the first transmission, often of 10\% \cite{google2019harq}. At the MAC layer, Hybrid Automatic Repeat reQuest (HARQ) uses incremental redundancy with Low-Density Parity-Check (LDPC) coding to retransmit additional parity bits when decoding fails, enabling partial forward error correction. When HARQ fails to recover the data within a limited number of attempts and the Radio Link Control (RLC) is operating in Acknowledged Mode, the RLC layer’s ARQ mechanism takes over by retransmitting entire packets. Unlike HARQ, RLC ARQ does not benefit from soft combining and introduces higher latency, making it less suitable for delay-sensitive applications \cite{adjakple2025user}. 

While HARQ offers faster recovery with moderate redundancy \cite{shen2009average} and ARQ ensures robustness at the expense of delay, both ultimately rely on feedback and retransmissions, which introduce round-trip time penalties. To improve the efficiency of such feedback-based systems, techniques have been explored such as optimizing block sizes for incremental redundancy rather than relying on brute-force MCS tuning \cite{heidarzadeh2018systematic}, modeling the delay impact of HARQ and ARQ by incorporating feedback and decoding latencies to estimate the probability that packets violate latency constraints under QoS guarantees \cite{moothedath2025delay}, or emphasizing more reliable detection of NACKs over ACKs to improve robustness in unreliable channels \cite{ding2021optimized}.

While feedback-based mechanisms like HARQ are well-established in current standards, their utility is increasingly being questioned in the context of stringent URLLC demands, given the overhead and delay they introduce \cite{love2008overview}. Other technologies are being explored in order to enable 3GPP systems to meet URLLC demands. These include higher-resolution beamforming, reconfigurable intelligent surfaces (RIS), and holographic radio (HR) at the physical layer \cite{pourkabirian2024vision}, tighter time and frequency synchronization and more fine-grained scheduling \cite{itur2022m2516}, and novel paradigms such as semantic communication which aims to reduce retransmission overhead while preserving meaning, thereby improving reliability under tight delay budgets \cite{adjakple2025user}.

A promising direction for addressing the limitations of HARQ and ARQ under URLLC constraints is the use of forward erasure correction, with network coding offering a flexible, non-feedback-dependent approach to achieving both reliability and low latency. Originally introduced as a theoretical framework for achieving multicast capacity in networks \cite{ahlswede2000network, ho2006random}, network coding has since evolved into a practical tool for end-to-end erasure correction and reliability enhancement. Random Linear Network Coding (RLNC), in particular, became a foundational technique for practical deployment, supporting dynamic coding decisions across variable channel conditions. Early studies, such as TCP/NC \cite{sundararajan2011tcpnc}, integrated a network coding layer between the transport and network layers, demonstrating throughput gains and robustness in lossy environments. Further work extended TCP/NC to bursty channels such as millimeter waves \cite{biyikoglu2025modeling}, a promising direction for 6G systems. Network coding’s performance relative to HARQ and ARQ was directly examined over WIMAX in \cite{teerapittayanon2012network}, with promising latency and reliability improvements. More recently, RLNC has been adapted for low-latency and throughput-sensitive use cases through techniques such as adjustable redundancy \cite{dilanchian2024adjustable}, adaptive causal coding \cite{cohen2020adaptive}, and practical implementation pipelines for variable conditions \cite{vasudevan2023practical}. The successful integration of RLNC into the 5G MAC layer, showing performance benefits as an alternative to HARQ and ARQ, was demonstrated in \cite{landon2024enhancing}, while a real-world deployment over the OpenAirInterface 5G testbed further validated its feasibility \cite{lhamo2024measurement}.

To support the URLLC demands in 5G and beyond, recent efforts have pushed network coding deeper into the architecture. Adaptive causal RLNC (AC-RLNC) has been developed to support multi-path and multi-hop scenarios, intelligently distributing redundancy across paths in a water-filling fashion to meet latency and reliability targets \cite{cohen2020adaptive}. Architectural studies have also explored how network coding can be orchestrated within Software Defined Networks (SDNs) to provide dynamic, centralized control over coding strategies in heterogeneous environments \cite{cohen2021bringing}. In device-to-device cellular systems, network coding has been shown to reduce packet completion time and improve spectrum efficiency \cite{keshtkarjahromi2018device}. Additional research has focused on customizing coding techniques for transport protocols, such as integrating FEC with QUIC to provide application-tailored reliability under tight timing constraints \cite{michel2022flec}. The limitations of traditional rateless codes in meeting latency guarantees have also prompted interest in sliding window RLNC approaches, which have been shown to satisfy the stringent constraints of millimeter-wave URLLC applications \cite{dias2023sliding}. Finally, coding schemes that multiplex streams with heterogeneous decoding deadlines, as explored in \cite{fong2020optimal}, are contributing to a growing toolbox of network coding strategies that offer structured, low-delay reliability guarantees in future wireless systems.

Over the course of the many refinements to network coding and its demonstrated potential for integration into real systems, most performance evaluations remain limited to simulations. There is a need for more testbed-based studies, particularly within the context of 3GPP-compliant mobile networks, to validate network coding under realistic channel and protocol conditions. This paper contributes to addressing that gap by conducting an experiment over a physical RF link between an Amarisoft UE simulator and callbox, providing practical insights into the viability of network coding in real-world 5G environments.

\section{Methodology}

This section outlines the design and implementation of an IP-layer RLNC module for real-time 5G traffic. Our goal is to enable forward erasure correction (FEC) without modifying the existing 5G protocol stack, using a modular netfilter-based interception and coding framework. The creation of a network coding layer which is invisible to the layers above and below has two components: logic to intercept packets from an application for modification and forward them on, and logic to implement RLNC over the packets that are intercepted. These modules operate transparently to applications and lower protocol layers.

\subsection{5G System Configuration}

We use an Amarisoft UE simulator (on Ubuntu 20.04.1 LTS, kernel version 5.15.0-139-generic) and callbox gNB simulator (on Fedora 34 (Workstation Edition), kernel version 5.11.12-300.fc34.x86\_64) as the two endpoints in which to implement network coding. Amarisoft’s AMARI UE Simbox and Callbox are software-defined user-equipment and eNodeB simulators which offer logging tools and WebSocket APIs for automated testing across full protocol stacks. In our setup, these devices communicate over a 3GPP-compliant RF link in a standalone 5G configuration using 20 MHz of bandwidth and Frequency Division Duplexing (FDD). 

We focus on the downlink scenario, where IP-layer packets are encoded at the gNB and decoded at the UE. When network coding is active, HARQ and ARQ are disabled by setting maximum HARQ transmissions to 1 and by setting the RLC layer to Unacknowledge Mode mode. To isolate the effects of erasure coding, the modulation and coding scheme (MCS) is fixed at index 14 (specifying a modulation order of 6 and LDPC code rate of 616/1024 as given in Table 5.1.3.2 of \cite{3gpp.38.214}) in all experiments to prevent link adaptation. This allows us to control the channel behavior via transmission power and observe direct relationships between signal-to-noise ratio (SNR), loss rate, and throughput under varying redundancy strategies.

\subsection{Packet Interception and Routing} \label{sec:nfq_interept}

We leverage the Linux netfilter subsystem to intercept IPv4 packets at two key points in the packet traversal path. Netfilter \cite{libnetfilterqueue} is a modular framework in the Linux kernel for intercepting, inspecting, and modifying packets as they traverse the networking stack. It introduces well-defined “hooks” in the protocol stack (e.g., for IPv4), where registered kernel modules can examine and act on packets by accepting, dropping, modifying, or queuing them to userspace via the NFQUEUE target. We used \texttt{libnetfilter\_queue} version 1.0.3 on the UE simulator and 1.0.2 on the callbox.

In our setup, we utilize the \texttt{NF\_IP\_POST\_ROUTING} hook (after routing is complete) on the sending side and the \texttt{NF\_IP\_PRE\_ROUTING} hook (before any routing is done) on the receiving side. At these two hooks, we register an \texttt{iptables} rule to route all packets to an \texttt{nfqueue}, where packets are made available to be manipulated in userspace.



\begin{figure}[!h]
    \centering
    \includegraphics[width=0.75\columnwidth]{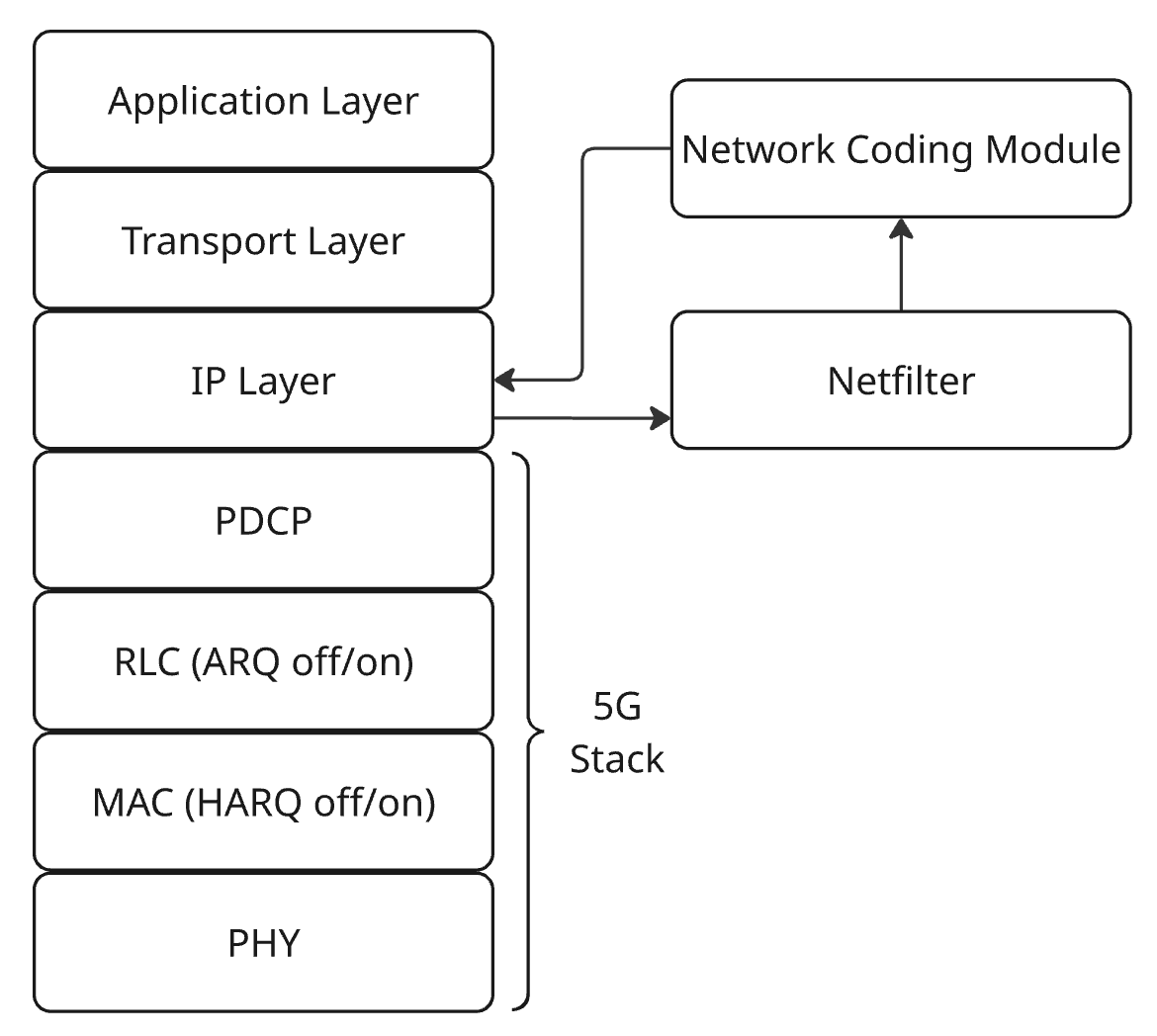}
    \caption{Placement of network coding in the networking stack.}
    \label{fig:nc_network_stack}
\end{figure}

We need to be able to intercept packets on a given port (from iperf or any other application), modify (i.e. encode) them, insert redundant packets as needed, and then place them back in the stream. This intercept program would run on the IP layers of the UE simulator and callbox gNB simulator, which communicate via a 3GPP RF link. This intercept program would later be combined with network coding, when HARQ and ARQ are deactivated on the gNB in order to replace retransmission-based redundancy with network coding. Fig. \ref{fig:nc_network_stack} illustrates the placement of network coding in the network stack in this work. Two additions are depicted: a netfilter module, and a network coding module. This separation is valuable because if a network coding layer were desired at a different point in the network coding stack, the network coding module could remain essentially unchanged; the only thing that would need to change is the method for intercepting packets.

\begin{figure*}
    \centering
    \includegraphics[width=0.75\textwidth]{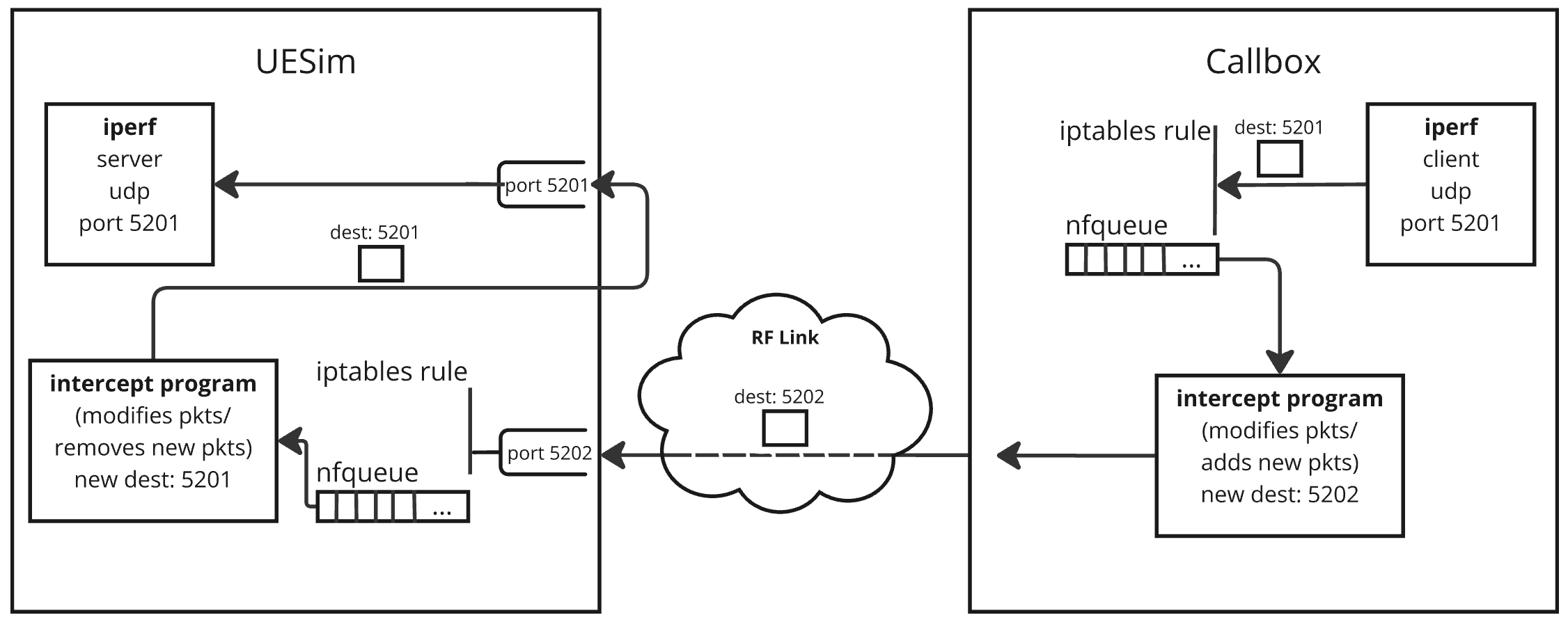}
    \caption{Packet interception system design.}
    \label{fig:nfq_sys_design}
\end{figure*}

Fig. \ref{fig:nfq_sys_design} illustrates how exactly we implemented this intercept program. An iptables rule is set to reroute packets from port 5201 (the iperf default port) to a netfilter queue. Packets in the queue are processed by an intercept program, which pops each packet, modifies it, and sends it to its original IP address but modifies the port to 5202. To avoid recursive interception during reinjection, encoded packets are forwarded via raw sockets to a secondary port (e.g., 5202), bypassing the \texttt{iptables} rule associated with the original port. The receiving node reverses this operation, decoding coded packets and forwarding the recovered payload to the original application port. The use of a raw socket ensures that the source and destination information are unchanged so that packet interception is invisible to higher layers,
as well as enabling the intercept program to periodically add new packets to the pipeline. The UE simulator has a similar iptables rule to route incoming packets on port 5202 to a netfilter queue. The receiving intercept program has the same capabilities as on the sending side. After modifying or removing packets from the queue, it sends them to its own IP address on port 5201, where the iperf server is listening. We used \texttt{iperf 2.1.6} and sent UDP traffic.

The implementation of the netfilter packet interception system design as laid out in section \ref{sec:nfq_interept} runs with no detectable impact on the jitter, throughput, or latency reported by iperf. Network coding logic was then added to the intercept program and collected data to compare network coding to existing systems. Note that in this paper, our evaluation focused on iperf-generated UDP traffic to quantify basic throughput and jitter metrics. However, the netfilter interception system is agnostic to the data being transferred other than the outgoing and incoming ports, meaning that this system can be adapted to test RLNC’s impact on application-layer performance such as video streaming for further exploration.

\subsection{RLNC Encoding Module}

The most straightforward implementation of random linear network coding (RLNC) is the classical block coding scheme. Certain modifications to the design of network coding can improve efficiency and latency, such as in the implementation known as sliding window coding\cite{karafillis2013algorithm} or adaptive coding\cite{cohen2020adaptive}. However, for this work, we stick to the block coding approach to provide an initial evaluation.

\subsubsection{Block Coding} Classical RLNC operates by sending $K$ original packets accompanied by $N-K$ redundant packets for a block size of $N$ total packets. In this scenario, each of the $N-K$ redundant packets, $R_i$, in the block is a linear combination of the $K$ original packets of the form:
\[
  R_{i}=\sum _{j=1}^{K}\rho _{i,j}P_{j} 
\]
where $\rho_{i,j}$ represents coefficients drawn from a sufficiently large field and $P_j$ represents original packet $j$ in the block. Classical RLNC is relatively easy to conceptualize and implement because it operates in discrete chunks, but it can introduce latency because encoded packets in a block cannot be decoded until $K$ total packets have been received \cite{dias2023sliding}.

\subsubsection{Sliding Window} Sliding window is the alternative to block-style network coding. Rather than dividing the outgoing data stream into discrete blocks of packets, it maintains a window of packets over which it calculates encoded packets. As new packets are sent, the window widens. As new packets are received and acknowledged, the window shrinks. Redundant packets are encoded as:
\[
R_{i}=\sum _{j=w_{\min }}^{w_{\max }}\rho _{i,j}P_{j}
\]
where $\rho_{i,j}$ represents coefficients drawn from a sufficiently large field and $P_j$ represents original packet $j$ in the window \cite{dias2023sliding}. This method reduces latency because the window size adapts to the needs of the network, allowing the system to decode redundant packets as it goes rather than waiting for the end of a block. An adaptive and causal RLNC in \cite{cohen2020adaptive} is another approach that provides a trade-off between latency and throughput requirements.

Encoding and decoding libraries for both block- and sliding window-style forms of network coding, KODO, are maintained and offered by the company Steinwurf \cite{pedersen2011kodo}. Steinwurf provided research access to these libraries in our implementations of RLNC on the JMA devices. This work implements and evaluates block RLNC, but leaves the implementation and evaluation of a sliding window coding module as future work.

\section{Results}
This section presents a quantitative comparison between conventional retransmission mechanisms (HARQ/ARQ) and RLNC-based forward erasure correction under varying channel conditions. Metrics of interest include throughput, jitter, and loss rate as measured using iperf over a 5G RF link against varying transmission gain that emulates channel degradation.

\subsection{Baseline: HARQ/ARQ Performance}

\begin{figure}[!h]
    \centering
    \includegraphics[width=\columnwidth]{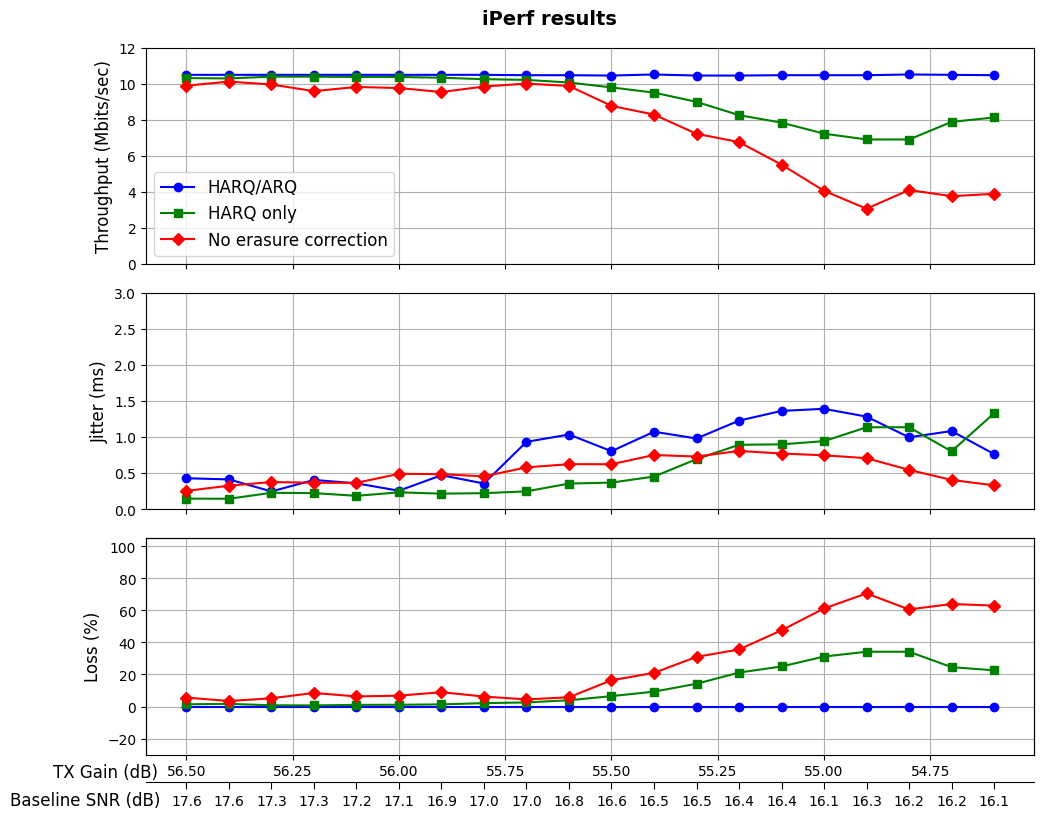}
    \caption{Throughput, jitter, and loss rates of HARQ/ARQ, HARQ only, and no erasure correction at different transmission gain settings.}
    \label{fig:no_nc_results}
\end{figure}

The implementation of the netfilter packet interception system design, as outlined in Section~\ref{sec:nfq_interept}, introduces no measurable overhead to jitter, throughput, or latency based on iperf evaluation. We first characterize baseline performance using traditional reliability schemes.

Fig. \ref{fig:no_nc_results} displays iperf metrics for 10 Mbps traffic between the callbox gNB simulator and UE simulator at various transmission gain settings. Transmission gain was swept from $56.5$ dB to $54.5$ dB in $0.1$ dB increments, producing a corresponding range of SNR values from $17.6$ to $16.1$ dB. This setup emulates channel degradation and allows for an understanding of the performance of RLNC at various SNR.

The curve labeled ``no erasure correction" presents a baseline reference. As expected, throughput achieves the requested rate at high gain, but drops quickly as gain decreases. Corresponding loss rises, with some jitter caused by irregular reception delays.

The curve labeled ``HARQ/ARQ" shows consistently high throughput and zero loss, but with higher fluctuation in the jitter early on. This suggests that, as expected, while ARQ achieves error recovery, it contributes significantly to end-to-end timing variation.

The ``HARQ only" curve helps to isolate the contribution of each reliability mechanism. The fact that the jitter early on is somewhere between that of HARQ/ARQ and no erasure correction demonstrates that a significant amount of the jitter is caused by ARQ. This is unsurprising given that ARQ essentially resets a packet's HARQ retransmission, and does so up to 8 times per packet, significantly increasing the potential delay per packet in exchange for more opportunities to receive it correctly. The loss rate is also interesting to compare to that of no correction, because for positive loss rates, HARQ halves the loss rate and ARQ brings that half rate down to zero.

\subsection{RLNC}

\begin{figure}[!h]
    \centering
    \includegraphics[width=\columnwidth]{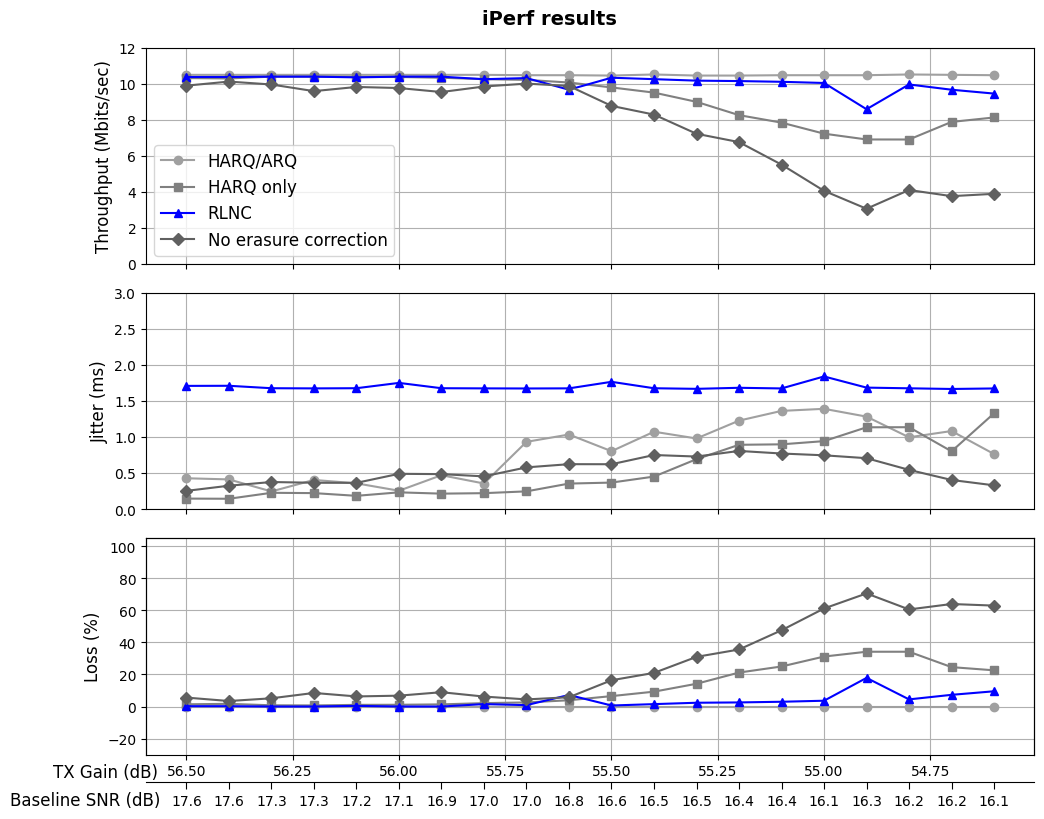}
    \caption{Throughput, jitter, and loss rates of block network coding at a 2/3 code rate (block size of 10 original packets with 5 redundant packets), compared with previously shown systems at different transmission gain settings.}
    \label{fig:nc_v_all_results}
\end{figure}

Figure~\ref{fig:nc_v_all_results} shows the performance of block RLNC at a fixed code rate of 2/3. This code rate can handle up to 33\% loss rates, seen in the plot at a transmission gain of about 55.3 dB. As shown, throughput remains stable and loss is near zero until the channel loss rate exceeds the protection capacity.
Unlike HARQ/ARQ, the jitter remains high but consistent across conditions. This is attributed to the block decoding delay inherent to classical RLNC—packets are decoded only after $K$ original packets are received. This fixed decoding schedule imposes a constant jitter penalty that is independent of network-induced variability. For applications tolerant to predictable delays, this tradeoff may be acceptable.

\begin{figure}[!h]
    \centering
    \includegraphics[width=\columnwidth]{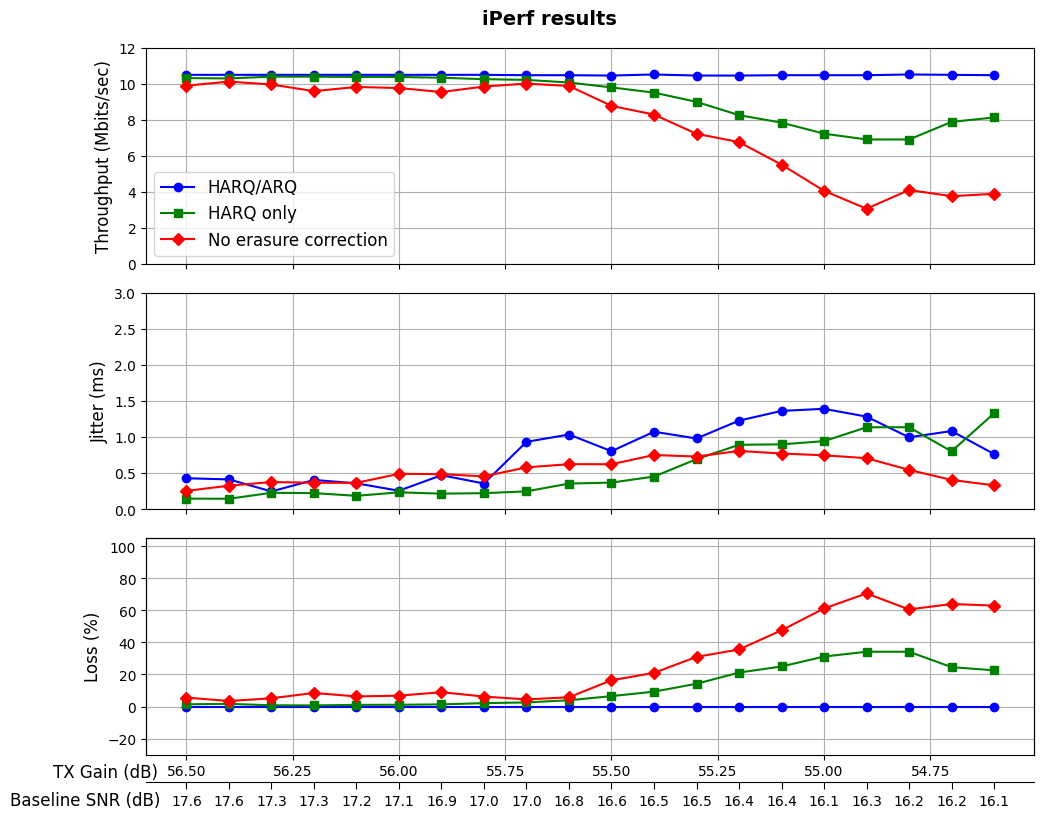}
    \caption{Throughput, jitter, and loss rates of HARQ/ARQ, HARQ only, and no erasure correction at different transmission gain settings.}
    \label{fig:no_nc_results}
\end{figure}

\begin{figure}[!h]
    \centering
    \includegraphics[width=\columnwidth]{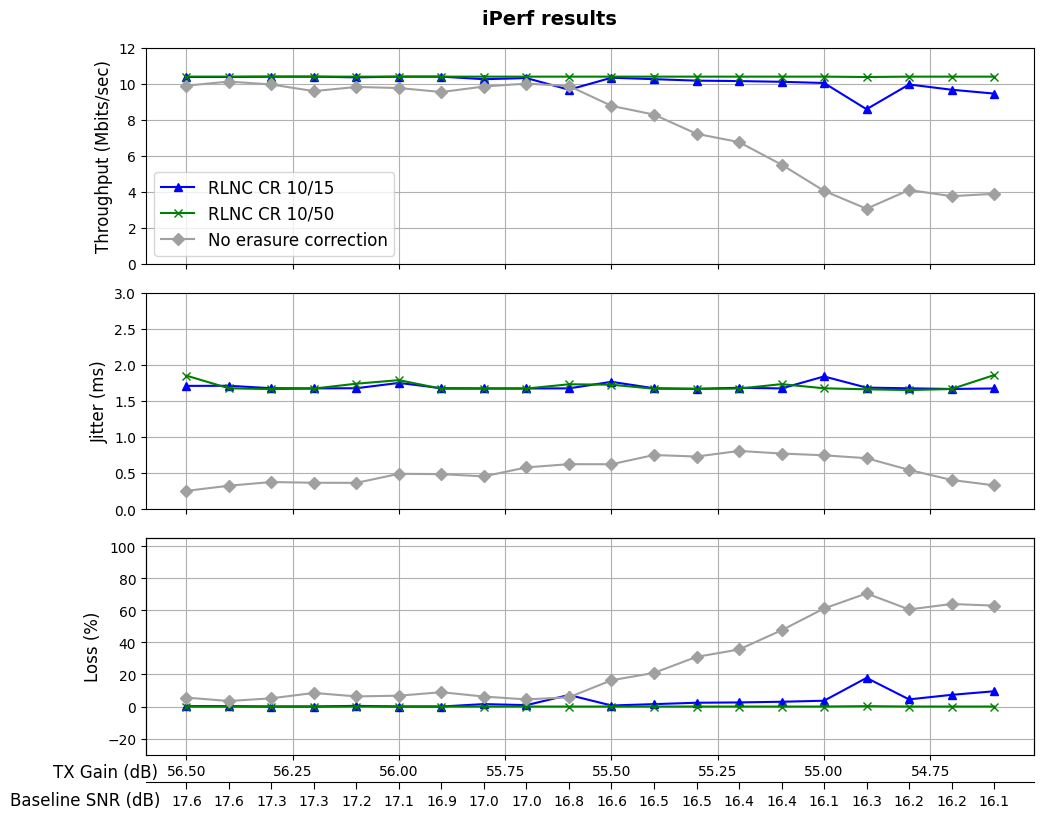}
    \caption{Throughput, jitter, and loss rates of block network coding at 2/3 vs. 1/5 code rates (block sizes of 10 original packets with 5 or 40 redundant packets, respectively) at different transmission gain settings.}
    \label{fig:dif_CRs}
\end{figure}

These results showcase the performance of a fixed code rate of 2/3 against a fixed MCS approach. With the fixed code rate, when the channel quality goes below the protected range, the performance of network coding system declines. This is also applicable to the fixed MCS scheme, but in practical 5G systems, an adaptive choice of MCS against link quality is utilized. This link adaptation scheme or an adaptive MCS approach selects a lower MCS value when the link quality deteriorates, ie, when it experiences losses. To match the adaptive MCS scheme, network coding can also use an approach to change the code rate based on the channel quality. While the prior test fixed the code rate to 2/3, a fully realized real-time implementation of network coding would adaptively change the code rate to match the current estimated channel conditions. RLNC is a form of forward erasure correction, which relies on estimation of the error rate in order to select an appropriate code rate. Fig. \ref{fig:dif_CRs} demonstrates the difference in loss between two different code rates. 

In this figure, a code rate of 2/3 and a code rate of 1/5 have similar behavior at low to moderate loss levels. As channel degradation increases, the code rate of 1/5 is able to maintain a near zero loss rate while the system using 2/3 code rate begins to fail. This low loss rate comes at the cost of a high redundancy rate, because an additional 4 times the target throughput (10 Mbps) must be sent in redundant packets. Thus at high loss rates a system would use a low code rate, and at low loss rates a system should transition to a higher code rate to achieve a better effective throughput. In a practical deployment, a dynamic RLNC scheme could estimate real-time channel loss and adjust the code rate accordingly to balance redundancy and resource efficiency.



\section{Discussion}
A correctly chosen RLNC code rate (one which is suitable for the current error rate) is able to correct losses and ensure full throughput. Existing HARQ/ARQ systems can also maintain throughput, but what the iperf throughput value alone does not reveal the gains RLNC wins in terms of resource usage. This comparison is not directly visible in plots produced by the current system because network coding is running on a higher layer than HARQ/ARQ. Our IP-layer design provides a modular starting point toward the integration of RLNC into 3GPP, with a longer-term goal of evaluating MAC-layer RLNC against native HARQ under equivalent scheduling conditions in order to more directly see resource usage comparisons. For this work, this comparison is demonstrated by analysis. This section provides an analysis of the resources used in our system and shows a clear advantage for network coding.

\subsection{Resource Usage Analysis}

We quantify the redundancy overhead associated with HARQ/ARQ and RLNC by modeling the expected number of transmissions per successfully delivered packet. As pointed out in the discussion of Fig. \ref{fig:no_nc_results}, empirical results indicate that approximately $50\%$ of the un-corrected loss is corrected by HARQ only and the other $50\%$ of the un-corrected loss is brought down to zero by combining with ARQ. Building on this, we estimate that the number of packets (counting each retransmission) which must be sent per original source packet is a function of the probability of each correction method succeeding and the number of packets each correction method requires. For instance, if we consider the loss rate $20\%$, which is within the range for which the $2/3$ code rate is sufficient (i.e. loss rates below $1/3$), then the total number of packets needed per source packet needed to reach zero loss for HARQ/ARQ can be estimated as 

\begin{equation*}
\begin{split}
    p_{HA,min} &= 
    \underset{\text{Successful in first Tx}}{(1 \times 0.8)} 
    + \underset{\text{after first HARQ RTx}}{(2 \times 0.1)} \\
    & + \underset{\text{after first ARQ RTx}}{(6 \times 0.1)} = 1.6
\end{split}
\end{equation*}
and 

\begin{equation*}
\begin{split}
    p_{HA,max} &= 
    \underset{\text{Successful in first Tx}}{(1 \times 0.8)} 
    + \underset{\text{after last HARQ RTx}}{(5 \times 0.1)} \\
    & + \underset{\text{after last ARQ round}}{(40 \times 0.1)} = 5.3
\end{split}
\end{equation*}

where $p_{HA,min}, p_{HA,max}$ represent the minimum and maximum total number of packets needed per source packet to reach zero loss for HARQ/ARQ. For the corresponding $2/3$ code rate, network coding sends 
\begin{align*}
   p_{NC} = 1/\texttt{CR} =  1.5 
\end{align*}
packets per source packet to reach zero loss. (Note that this is a simplified analysis - the network coding code rate here overestimates the loss rate significantly, and could be adjusted to more closely match the exact amount of loss.)

This calculation was performed for a $20\%$ loss rate, but this can be abstracted to apply to any loss rate $r$. 

\begin{align*}
p_{HA,min} &= 
(1 \times (1-r))
+ (2 \times r/2) 
+ (6 \times r/2) \\
&= 1+3r
\end{align*}
and

\begin{align*}
p_{HA,min} &= 
(1 \times (1-r))
+ (5 \times r/2) 
+ (40 \times r/2) \\
&=  1+21.5r
\end{align*}
and 

\begin{align*}
    p_{NC} &\geq \frac{1}{1-r}.
\end{align*}

For all loss rates $0 \leq r \leq 2/3$, we see that $p_{NC} \leq p{HA, min}$. 5G systems are often run in such a way as to target a $10\%$ loss rate, which would result in a need for $1.11$ network coded packets per source packet, or between $1.3$ and $3.15$ packets per source packet using HARQ/ARQ. Thus network coding offers better use of resources in a bandwidth-limited system.

In the earlier study performed over WIMAX \cite{teerapittayanon2012network}, similar tests were done and it was found that network coding achieved up to 5.9 times the throughput attainable by HARQ/ARQ. The study was performed at approximately $20\%$ loss rate, which in this analysis would require approximately $1.25$ network coded packets per source packet, or between $1.6$ and $5.3$ packets per source packet using HARQ/ARQ, or up to a $4.24$ throughput improvement for network coding. Thus, similar results are achievable for 5G as were demonstrated for WIMAX.

These estimates demonstrate that RLNC can offer lower average redundancy per delivered packet, particularly in moderate loss regimes. This efficiency gain makes RLNC attractive in bandwidth-constrained deployments.

\subsection{Reducing Jitter}
We analyze how decoding strategies affect packet inter-arrival consistency and explore techniques for minimizing jitter introduced by block decoding. As mentioned previously, the jitter measured by iperf in Fig. \ref{fig:nc_v_all_results} shows a higher value for RLNC than for the other systems. This jitter is caused by the blocks in block coding. 

The sending intercept network coding program receives packets at whatever spacing is set by iperf, generates a coded version of that packet according to the current block, saves the current packet information in memory, and sends that packet immediately. After $k$ packets, $n-k$ redundant packets are sent in a burst. At the decoder, however, packets are not decoded until $k$ packets of a given block are received, at which point $k$ packets are decoded in a burst. This results in a relatively stable jitter value independent of loss rate because the limiting factor is the length of the block.

Because this jitter effect is predictable and caused by a cyclic buildup and burst of packets, there may be jitter-sensitive applications for which this jitter effect is not problematic. Network coding ensures that delay or rearrangement of individual packets is smoothed out by the decoding process, which means that if an application can accept a delay offset of at least the length of a single block, the queueing rate of packets is very reliable and unlikely to overflow or underflow a packet buffer. Real-time applications with buffering, such as video streaming or VoIP, can leverage this determinism by adjusting their playback buffer to match the block interval. Unlike feedback-based systems, which may produce unpredictable packet spacing due to variable retransmissions, block RLNC’s burst pattern enables tighter control over playback timing and resource provisioning.

In order to address the needs of applications that have strictly defined low jitter requirements, an on-the-fly decoding process can be implemented. This can be introduced by using systematic packets with a combination of coded packets in a block and/or by using sliding window network coding. Another technique that can reduce the packet jitter is to use a ``drop when seen" acknowledgment instead of a ``drop when decoded" approach. The drop when decoded approach is used in block decoding, where packet jitter is measured from the moment the packets in a block are decoded, which happens in a burst. In the drop when seen approach, the receiver acknowledges every new degree of freedom of information rather than the original packet \cite{sundararajan2008arq}. This allows for earlier acknowledgment of packets, and thus reduces the jitter. Preliminary efforts toward integrating sliding window decoding in this system have begun, and a detailed evaluation is planned as future work.

\section{Future Work}

Building on our preliminary efforts to integrate sliding window decoding, future work will focus on fully deploying this technique to enable per-packet decoding and reduce block-induced jitter. This improvement would better support latency-sensitive traffic and align with the requirements of URLLC use cases. Beyond decoding strategy, integrating adaptive rate control and channel feedback would enable the encoder to dynamically adjust code rates based on real-time loss estimates, akin to how modern 5G systems adapt modulation and coding schemes. Combining these changes with future experiments which incorporate traffic profiles representative of real applications, including video streaming and low-latency control flows, will help to better understand how RLNC interacts with higher-layer protocol dynamics. These enhancements would elevate our current RLNC prototype from a proof-of-concept into a deployable alternative to HARQ and ARQ, with the responsiveness and efficiency necessary for next-generation wireless systems.

Future work should also focus on integrating network coding deeper into the 5G protocol stack. As demonstrated in this paper, the network coding module is modular and could be deployed at different layers, provided appropriate packet interception is available. Moving the implementation to the MAC or RLC layer would allow for more direct comparisons with existing feedback-based mechanisms and could facilitate standardization efforts. Positioned correctly within the stack, network coding has the potential to transform how wireless networks manage the tradeoff between latency, reliability, and spectrum efficiency.

In the long term, network coding research must explore the path toward integration into standardized 5G and 6G protocol stacks. Embedding network coding functionality into lower layers such as MAC or RLC, where tighter control over scheduling and retransmission decisions can further reduce latency and overhead, will provide the best results. Realizing this vision would require standardization efforts within 3GPP, potentially involving extensions to current RLC modes or the development of coding-aware variables in the scheduler. These changes must consider compatibility with existing HARQ processes and 3GPP stack-specific technologies. Similar to how HARQ is managed through logical channels and process IDs, network-coded packets may require dedicated logical channels or new signaling procedures. Moreover, integration into the SDN or NFV control plane could enable centralized rate adaptation, coding strategy selection, and cross-layer optimization \cite{cohen2021bringing}. By framing RLNC not as a replacement but as a configurable reliability option in the protocol stack, there is potential to incrementally introduce it into commercial deployments without disrupting existing systems.

\section{Conclusion}

We presented a practical implementation of Random Linear Network Coding (RLNC) at the IP layer of a 5G system, enabling forward erasure correction as an alternative to retransmission-based mechanisms like ARQ and HARQ. Using netfilter for real-time packet interception, we demonstrated that RLNC can fully compensate for moderate to high packet loss while requiring fewer redundant transmissions than traditional feedback-based approaches. Our block coding implementation successfully maintained full throughput under adverse conditions where retransmissions would incur significant latency.

Although the decoding delay introduced jitter, this tradeoff was shown to be regular and predictable, potentially acceptable for many streaming and buffered applications. This system provides valuable insight into the behavior and potential of network coding in real-world 5G deployments and serves as a step toward more efficient and flexible reliability mechanisms aligned with 3GPP goals.

\bibliographystyle{IEEEtran}
\bibliography{IEEEabrv,references}

\begin{thebibliography}{10}
\providecommand{\url}[1]{#1}
\csname url@samestyle\endcsname
\providecommand{\newblock}{\relax}
\providecommand{\bibinfo}[2]{#2}
\providecommand{\BIBentrySTDinterwordspacing}{\spaceskip=0pt\relax}
\providecommand{\BIBentryALTinterwordstretchfactor}{4}
\providecommand{\BIBentryALTinterwordspacing}{\spaceskip=\fontdimen2\font plus
\BIBentryALTinterwordstretchfactor\fontdimen3\font minus
  \fontdimen4\font\relax}
\providecommand{\BIBforeignlanguage}[2]{{%
\expandafter\ifx\csname l@#1\endcsname\relax
\typeout{** WARNING: IEEEtran.bst: No hyphenation pattern has been}%
\typeout{** loaded for the language `#1'. Using the pattern for}%
\typeout{** the default language instead.}%
\else
\language=\csname l@#1\endcsname
\fi
#2}}
\providecommand{\BIBdecl}{\relax}
\BIBdecl

\bibitem{wp5d2023m}
I.-R. Report, ``M. 2160: Framework and overall objectives of the future
  development of imt for 2030 and beyond,'' \emph{International
  Telecommunication Union, Report M}, 2023.

\bibitem{ho2006random}
T.~Ho, M.~M{\'e}dard, R.~Koetter, D.~R. Karger, M.~Effros, J.~Shi, and
  B.~Leong, ``A random linear network coding approach to multicast,''
  \emph{IEEE Transactions on information theory}, vol.~52, no.~10, pp.
  4413--4430, 2006.

\bibitem{landon2024enhancing}
L.~Landon, V.~A. Vasudevan, J.~Kim, J.~Sung, J.~T. Masters, and M.~M{\'e}dard,
  ``Enhancing 5g performance: Reducing service time and research directions for
  6g standards,'' in \emph{2024 3rd International Conference on 6G Networking
  (6GNet)}.\hskip 1em plus 0.5em minus 0.4em\relax IEEE, 2024, pp. 37--45.

\bibitem{teerapittayanon2012network}
S.~Teerapittayanon, K.~Fouli, M.~M{\'e}dard, M.-J. Montpetit, X.~Shi,
  I.~Seskar, and A.~Gosain, ``Network coding as a wimax link reliability
  mechanism,'' in \emph{Multiple Access Communications: 5th International
  Workshop, MACOM 2012, Maynooth, Ireland, November 19-20, 2012. Proceedings
  5}.\hskip 1em plus 0.5em minus 0.4em\relax Springer, 2012, pp. 1--12.

\bibitem{alves2021beyond}
H.~Alves, G.~D. Jo, J.~Shin, C.~Yeh, N.~H. Mahmood, C.~Lima, C.~Yoon,
  N.~Rahatheva, O.-S. Park, S.~Kim \emph{et~al.}, ``Beyond 5g urllc evolution:
  New service modes and practical considerations,'' \emph{arXiv preprint
  arXiv:2106.11825}, vol.~7, 2021.

\bibitem{pourkabirian2024vision}
A.~Pourkabirian, M.~S. Kordafshari, A.~Jindal, and M.~H. Anisi, ``{A vision of
  6G URLLC: Physical-layer technologies and enablers},'' \emph{IEEE
  Communications Standards Magazine}, vol.~8, no.~2, pp. 20--27, 2024.

\bibitem{google2019harq}
G.~LLC, ``Link adaptation in wireless communication systems,'' World
  Intellectual Property Organization (WIPO), WO2019155249A1, 2019, [Online].
  Available: \url{https://patents.google.com/patent/WO2019155249A1}.

\bibitem{adjakple2025user}
P.~Adjakple, A.~Ijaz, J.~Cray, A.~Almradi, J.~Huang, and D.~Castor, ``User
  plane design approaches for 6g,'' \emph{IEEE Wireless Communications},
  vol.~32, no.~3, pp. 8--11, 2025.

\bibitem{shen2009average}
C.~Shen, T.~Liu, and M.~P. Fitz, ``On the average rate performance of
  hybrid-arq in quasi-static fading channels,'' \emph{IEEE Transactions on
  Communications}, vol.~57, no.~11, pp. 3339--3352, 2009.

\bibitem{heidarzadeh2018systematic}
A.~Heidarzadeh, J.-F. Chamberland, R.~D. Wesel, and P.~Parag, ``A systematic
  approach to incremental redundancy with application to erasure channels,''
  \emph{IEEE Transactions on Communications}, vol.~67, no.~4, pp. 2620--2631,
  2018.

\bibitem{moothedath2025delay}
V.~N. Moothedath, S.~Seo, N.~Petreska, B.~Kloiber, and J.~Gross, ``Delay
  analysis of 5g harq in the presence of decoding and feedback latencies,''
  \emph{arXiv preprint arXiv:2502.08789}, 2025.

\bibitem{ding2021optimized}
W.~Ding and M.~Shikh-Bahaei, ``Optimized asymmetric feedback detection for
  rate-adaptive harq with unreliable feedback,'' in \emph{2021 IEEE Wireless
  Communications and Networking Conference (WCNC)}.\hskip 1em plus 0.5em minus
  0.4em\relax IEEE, 2021, pp. 1--6.

\bibitem{love2008overview}
D.~J. Love, R.~W. Heath, V.~K. Lau, D.~Gesbert, B.~D. Rao, and M.~Andrews, ``An
  overview of limited feedback in wireless communication systems,'' \emph{IEEE
  Journal on selected areas in Communications}, vol.~26, no.~8, pp. 1341--1365,
  2008.

\bibitem{itur2022m2516}
{International Telecommunication Union Radiocommunication Sector (ITU-R)},
  ``{Future Technology Trends of Terrestrial International Mobile
  Telecommunications Systems Towards 2030 and Beyond},'' ITU-R, Tech. Rep.
  M.2516-0, February 2022, [Online]. Available:
  \url{https://www.itu.int/rec/R-REC-M.2516-0-202202-I/en}.

\bibitem{ahlswede2000network}
R.~Ahlswede, N.~Cai, S.-Y. Li, and R.~W. Yeung, ``Network information flow,''
  \emph{IEEE Transactions on information theory}, vol.~46, no.~4, pp.
  1204--1216, 2000.

\bibitem{sundararajan2011tcpnc}
J.~K. Sundararajan, D.~Shah, M.~Médard, S.~Jakubczak, M.~Mitzenmacher, and
  J.~Barros, ``Network coding meets tcp: Theory and implementation,''
  \emph{Proceedings of the IEEE}, vol.~99, no.~3, pp. 490--512, 2011.

\bibitem{biyikoglu2025modeling}
Y.~A. Biyikoglu, V.~A. Vasudevan, and M.~M{\'e}dard, ``Modeling network
  coding-enabled protocols on bursty channels,'' in \emph{2025 59th Annual
  Conference on Information Sciences and Systems (CISS)}.\hskip 1em plus 0.5em
  minus 0.4em\relax IEEE, 2025, pp. 1--6.

\bibitem{dilanchian2024adjustable}
R.~Dilanchian, A.~Bohlooli, and K.~Jamshidi, ``Adjustable random linear network
  coding (arlnc): a solution for data transmission in dynamic iot computational
  environments,'' \emph{Digital Communications and Networks}, 2024.

\bibitem{cohen2020adaptive}
A.~Cohen, D.~Malak, V.~B. Bracha, and M.~M{\'e}dard, ``Adaptive causal network
  coding with feedback,'' \emph{IEEE Transactions on Communications}, vol.~68,
  no.~7, pp. 4325--4341, 2020.

\bibitem{vasudevan2023practical}
V.~A. Vasudevan, T.~Soni, and M.~M{\'e}dard, ``Practical sliding window
  recoder: Design, analysis, and usecases,'' in \emph{2023 IEEE 29th
  International Symposium on Local and Metropolitan Area Networks
  (LANMAN)}.\hskip 1em plus 0.5em minus 0.4em\relax IEEE, 2023, pp. 1--6.

\bibitem{lhamo2024measurement}
O.~Lhamo, T.~V. Doan, E.~Tasdemir, M.~Attawna, G.~T. Nguyen, P.~Seeling,
  M.~Reisslein, and F.~H. Fitzek, ``Measurement study of programmable network
  coding in cloud-native 5g and beyond networks,'' \emph{arXiv preprint
  arXiv:2408.06115}, 2024.

\bibitem{cohen2021bringing}
A.~Cohen, H.~Esfahanizadeh, B.~Sousa, J.~P. Vilela, M.~Luis, D.~Raposo,
  F.~Michel, S.~Sargento, and M.~Medard, ``Bringing network coding into {SDN}:
  Architectural study for meshed heterogeneous communications,'' \emph{IEEE
  Communications Magazine}, vol.~59, no.~4, pp. 37--43, 2021.

\bibitem{keshtkarjahromi2018device}
Y.~Keshtkarjahromi, H.~Seferoglu, R.~Ansari, and A.~Khokhar, ``Device-to-device
  networking meets cellular via network coding,'' \emph{IEEE/ACM Transactions
  on Networking}, vol.~26, no.~1, pp. 370--383, 2018.

\bibitem{michel2022flec}
F.~Michel, A.~Cohen, D.~Malak, Q.~De~Coninck, M.~M{\'e}dard, and
  O.~Bonaventure, ``{FlEC}: Enhancing {QUIC} with application-tailored
  reliability mechanisms,'' \emph{IEEE/ACM Transactions on Networking}, 2022.

\bibitem{dias2023sliding}
E.~Dias, D.~Raposo, H.~Esfahanizadeh, A.~Cohen, T.~Ferreira, M.~Luís,
  S.~Sargento, and M.~Médard, ``Sliding window network coding enables next
  generation urllc millimeter-wave networks,'' \emph{IEEE Networking Letters},
  vol.~5, no.~3, pp. 159--163, 2023.

\bibitem{fong2020optimal}
S.~L. Fong, A.~Khisti, B.~Li, W.-T. Tan, X.~Zhu, and J.~Apostolopoulos,
  ``Optimal multiplexed erasure codes for streaming messages with different
  decoding delays,'' \emph{IEEE Transactions on Information Theory}, vol.~66,
  no.~7, pp. 4007--4018, 2020.

\bibitem{3gpp.38.214}
3GPP-TS.38.214, ``{Physical layer procedures for data},'' {3rd Generation
  Partnership Project (3GPP)}, Technical Specification (TS) 38.214, 09 2020,
  version 16.3.0.

\bibitem{libnetfilterqueue}
\BIBentryALTinterwordspacing
{Linux Foundation}. (2025) {The Netfilter Project}. Accessed: 2025-05-30.
  [Online]. Available:
  \url{\url{https://netfilter.org/projects/libnetfilter\_queue/doxygen/html/}}
\BIBentrySTDinterwordspacing

\bibitem{karafillis2013algorithm}
P.~Karafillis, K.~Fouli, A.~ParandehGheibi, and M.~M{\'e}dard, ``An algorithm
  for improving sliding window network coding in tcp,'' in \emph{2013 47th
  Annual Conference on Information Sciences and Systems (CISS)}.\hskip 1em plus
  0.5em minus 0.4em\relax IEEE, 2013, pp. 1--5.

\bibitem{pedersen2011kodo}
M.~V. Pedersen, J.~Heide, and F.~H. Fitzek, ``Kodo: An open and research
  oriented network coding library,'' in \emph{NETWORKING 2011 Workshops:
  International IFIP TC 6 Workshops, PE-CRN, NC-Pro, WCNS, and SUNSET 2011,
  Held at NETWORKING 2011, Valencia, Spain, May 13, 2011, Revised Selected
  Papers 10}.\hskip 1em plus 0.5em minus 0.4em\relax Springer, 2011, pp.
  145--152.

\bibitem{sundararajan2008arq}
J.~K. Sundararajan, D.~Shah, and M.~M{\'e}dard, ``Arq for network coding,'' in
  \emph{2008 IEEE International Symposium on Information Theory}.\hskip 1em
  plus 0.5em minus 0.4em\relax IEEE, 2008, pp. 1651--1655.

\end{thebibliography}

\end{document}